\documentstyle[ms,11pt]{egs} 

\input epsf

\printfigures              

\begin{document}
\title{Comparative Evolution of Jupiter and Saturn}

%
%
%
%
\author[1]{W. B. Hubbard}
\affil[1]{Lunar \& Planetary Lab., The Univ. of Arizona, Tucson, AZ 85721-0092 USA}
\author[2]{T. Guillot}
\affil[2]{Obs. de la C\^ote d'Azur, F-06304 Nice Cedex 4, France}
\author [3]{M. S. Marley}
\affil[3]{Dept. of Astronomy, New Mexico State Univ., Las Cruces, NM 88003-0001 USA}
\author[4]{A. Burrows}
\affil[4]{Dept. of Astronomy, The Univ. of Arizona, Tucson, AZ 85721 USA}
\author[1,6]{J. I. Lunine}
\author[5]{D. S. Saumon}
\affil[5]{Dept. of Physics \& Astronomy, Vanderbilt Univ., Nashville, TN 37235 USA}
\affil[6]{Reparto di Planetologia,
CNR -- Istituto di Astrofisica, 00133 Roma Italy}
\date{Manuscript version from 9 December 1998}

\journal{\PSS}       
%
%
\firstauthor{Hubbard}
\proofs{W. B. Hubbard\\Lunar \& Planetary Lab.\\Tucson, AZ 85721-0092}
\offsets{W. B. Hubbard\\Lunar \& Planetary Lab.\\Tucson, AZ 85721-0092}

\msnumber{ }
\received{XX   November 1998}
\accepted{        }

\maketitle

\begin{abstract}

We present evolutionary
sequences for Jupiter and Saturn, based on new nongray model atmospheres,
which take into account the evolution of the solar luminosity and
partitioning of dense components to deeper layers.  The results are
used to set limits on the extent to which
possible interior phase separation of hydrogen and helium may have
progressed in the two planets.  
When combined with static models constrained
by the gravity field, our evolutionary calculations constrain the helium
mass fraction in Jupiter to be between 0.20 and 0.27, relative to total
hydrogen and helium. This is in agreement with the Galileo determination.
The helium mass fraction in Saturn's atmosphere lies between 0.11 and 0.25,
higher than the Voyager determination. Based on the discrepancy between the
Galileo and Voyager results for Jupiter, and our models, we predict that
Cassini measurements will yield a higher atmospheric
helium mass fraction for Saturn
relative to the Voyager value.

\end{abstract}

\section{Introduction}
\label{sec:intro}


The thermal evolution of Jupiter and Saturn is a long-standing problem
which couples the transport properties of the atmosphere and interior
of Jupiter/Saturn, the thermodynamics and phase diagram of the deep
interior, and the effect of solar heating of the atmosphere.  New
perspectives on this problem have been provided by recent experiments
on the metallization of hydrogen at high pressures \citep{collins98},
new results for the composition of the jovian
atmosphere from the Galileo entry probe \citep{vonzahn98}, and
recent work by our research group on the evolution of extrasolar giant
planets and brown dwarfs \citep{burr97}. 

The thermal evolution of a giant planet with an isentropic or
near-isentropic interior temperature distribution is parametrized
in terms of a surface which relates the three variables $T_{\rm eff}$,
$T_{10}$, and $g$, where $T_{\rm eff}$ is the effective temperature
at which the planet radiates its internally-derived and converted
solar energy into space, $T_{10}$ is the temperature at 10 bars pressure,
which characterizes the isentrope in the outer layers of the planet
(note that if the atmosphere is radiative at 10 bars, $T_{10}$ represents
the temperature on the deeper isentrope extrapolated to 10 bars),
and $g$ is the surface gravity.
The surface is shown in Fig. 1.

\begin{figure*}

\epsfxsize=14cm \epsfbox{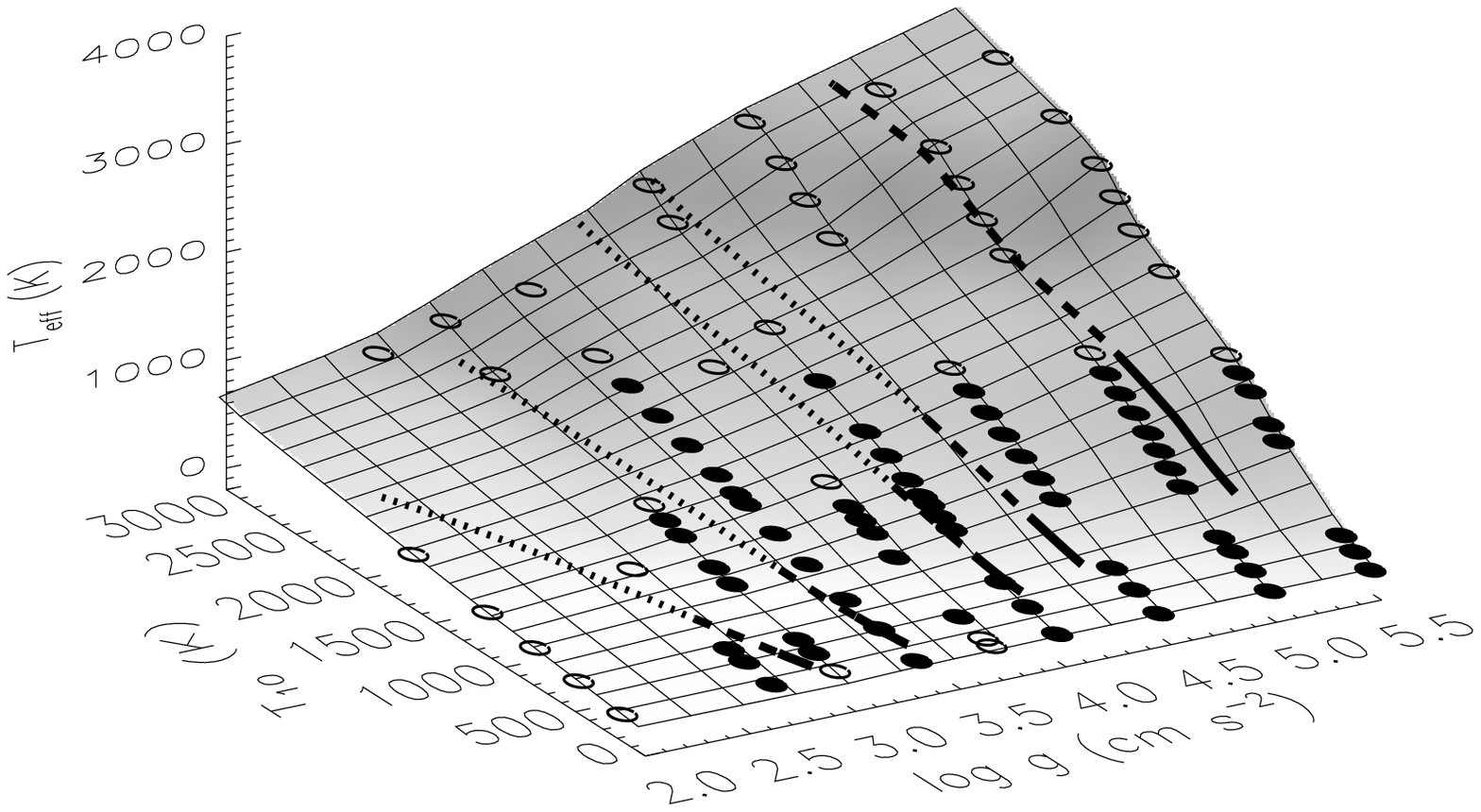}
  \caption[]{\label{fig:intro}
This surface is produced by a splined fit to individual model atmospheres
\citep{burr97}, augmented by additional unpublished calculations.  Solid
dots show non-gray calculations, while open circles show gray calculations.
The superimposed trajectories show evolution for homogeneous
isolated bodies with
masses equal to (lower left to upper
right) Saturn, Jupiter, 5 Jupiters, 10 Jupiters, and
42 Jupiters.  Dotted lines show evolution at age $t < 10^8$ years,
heavy dashed lines show evolution for $t$ between
$10^8$ and $10^9$ years, and solid lines show evolution at
$t > 10^9$ years.}
\end{figure*}

Figure 1 illustrates that the theory of the evolution of
Jupiter and Saturn can be subsumed within a larger study of giant
planets and brown dwarfs.  The principal difference between Jupiter
and Saturn and more massive bodies is that the evolution of the
former involves a possible phase separation of helium within
the metallic-hydrogen interior, as shown in Fig. 2 \citep{ss77}.

Evolution models of Jupiter and Saturn (Saumon et~al., 1992;
Guillot et~al., 1995) have so far assumed the planet to remain
homogeneous in the hydrogen-helium phase. This
is questionable in the case of Jupiter, and certainly wrong for
Saturn. 

\begin{figure*}

\epsfxsize=15cm \epsfbox{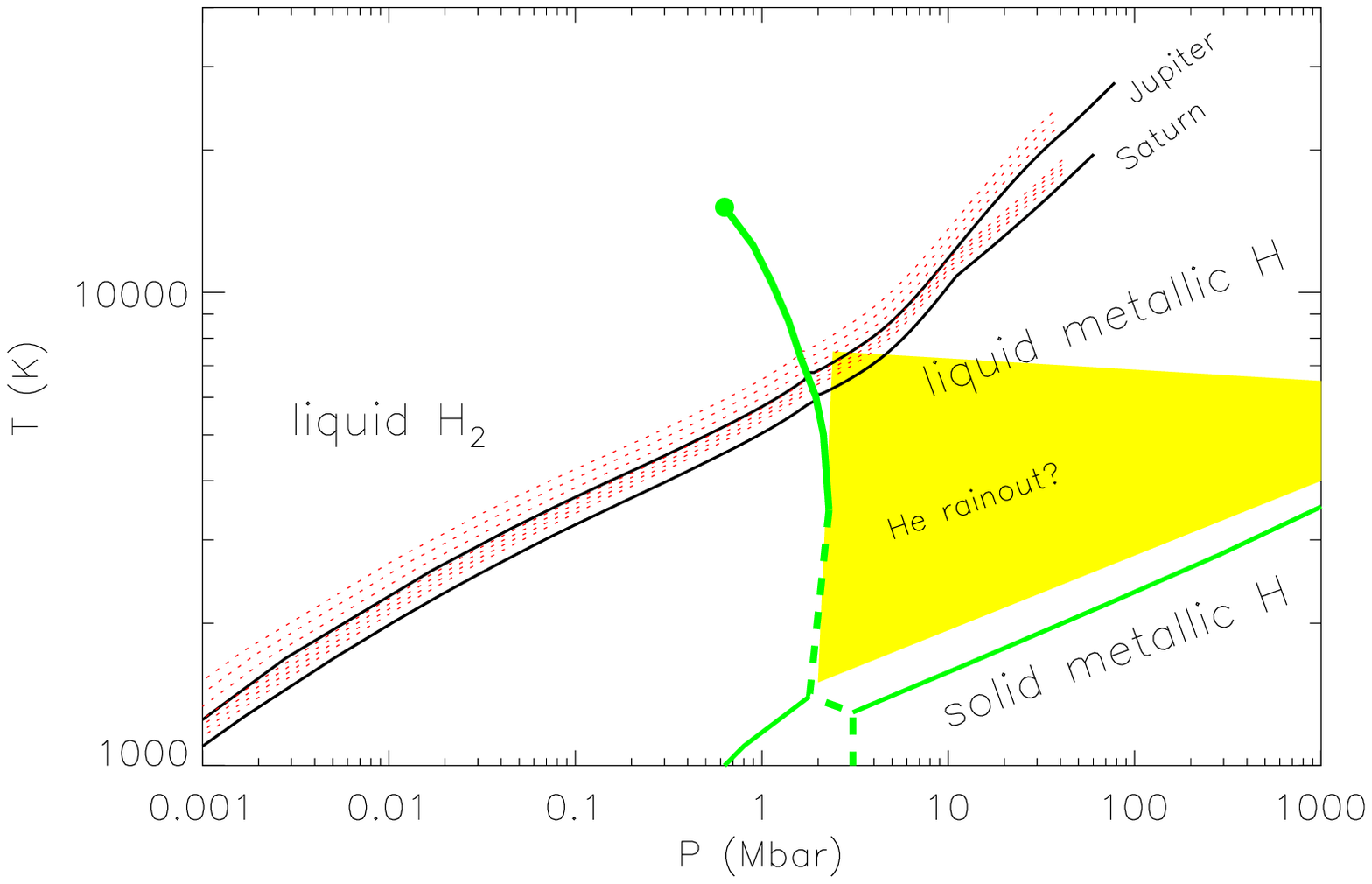}
  \caption[]{\label{figt:intro}
Phase diagram of hydrogen \citep{wbh97}, with evolution of Jupiter and Saturn.
Bodies more massive than Jupiter will not enter
the He rainout region within a Hubble time. The phase transition between
liquid molecular hydrogen and liquid metallic hydrogen is according
to the theory of \citet{scvh95}.}
\end{figure*}

In Fig. 2, the solid curves show present-day Jupiter and
Saturn, while the dashed curves show evolving models at
earlier epochs with homogeneous and
solar H-He proportions.  In a fully consistent treatment
of helium separation, the partitioning of helium in the
metallic-hydrogen core would be calculated in accordance with
a model free energy for a mixture of metallic hydrogen and helium.
This paper does not carry out such a treatment.  Rather,
we introduce two limiting versions of separation of helium
from hydrogen, carried out in such a way that the essential conclusions
can be applied to separation of any denser component (e.g.,
ice or rock) from hydrogen.

The helium distribution in a Jupiter or Saturn
model is characterized
by $Y$, the helium mass fraction of the hydrogen-helium
mixture in a given layer.  The hydrogen mass fraction
is then $X=1-Y$.
We characterize the value of $Y$ in initial models with
homogeneous hydrogen-helium composition by $Y_{\rm proto}=0.27$,
which we take to be the initial protosolar value.

In the first version of the theory,
discussed in Section 2, unmixing is assumed
to occur with a linear time dependence, with the denser
component (nominally He) being depleted by a constant
factor throughout the molecular envelope
and enriched by a constant factor throughout the metallic-hydrogen
interior. 
In the second version,
discussed in Section 3, unmixing occurs only in the final stages of
evolution of the planet,
occurring between
the last dashed-curve model and the solid-curve model of Fig. 2.
In the second version, the denser component is removed uniformly
from all hydrogen-rich layers and added to a dense core at the center
of the planet.

We will argue in the conclusion that the two versions of
chemical evolution set
limits on the possible unmixing in the two planets, and hence, on
the maximum helium depletion to be expected in the atmosphere.

\section{A semi-analytical non-homogeneous evolution model}

\subsection{Derivation}

We present an estimation of the delay in the cooling of the planets 
introduced by helium (or any other element) differentiation using a
simple analytical model. 

The following assumptions are used: (i) the planet is fully
isentropic, has no core, and is solely made of hydrogen and
helium. (ii) It is initially homogeneous, but a phase separation (or
phase transition) occurs at a fractional mass $m_{\rm t}(t)$ (note that
$m$ ranges from $0$ at the center to $1$ at the surface), which leads to the 
creation of an upper helium poor region (labelled I), and a deeper
helium-rich region (labelled II). The compositional difference between
the two regions is defined by the difference in their hydrogen mass
fraction $\Delta X(t)=X_{\rm I}-X_{\rm II}$. Note that $\Delta X(t)$
is a monotonically increasing function of $t$ and that $\Delta
X(0)=0$. 
(iii) The transition region between I and II is infinitely small, and
any change in composition due either to the variations of $m_{\rm t}$ or of
$\Delta X$ with time is instantaneously redistributed by convection,
so that regions I and II remain homogeneous and adiabatic. 
(iv) Finally, we assume that the energy released by the falling
helium droplets all goes into intrinsic planetary
luminosity. In fact, a small fraction (about 10\% perhaps) is retained
in the form of a higher internal temperature \citep{ss77}.

Following Hubbard's (1977) procedure, we derive the evolution time
scale from the energy conservation equation, but splitting the time
derivative of the specific entropy $S$ in two parts: a homogeneous,
and an inhomogeneous part:
\begin{equation}
{L\over M}=\int-T\left[\left({\partial S\over \partial t}\right)_X +{dX\over
dt}\left({\partial S\over\partial X}\right)_t\right]dm.
\label{eq:energy}
\end{equation}
Here $m$ is the mass fraction variable, normalized to unity at the
planet's surface.

We assume that the total specific entropy depends linearly on the
mass mixing ratios of hydrogen and helium, respectively
(thereby neglecting the small contribution due to the mixing entropy).
This yields
\begin{equation}
\left({\partial S\over \partial X}\right)_t=S_{\rm H}-S_{\rm He}
\equiv \delta S,
\label{eq:ds}
\end{equation}
where $S_{\rm H}$ and $S_{\rm He}$ are the specific entropies of pure
hydrogen and pure helium, respectively. 

Furthermore, using mass conservation in the planet between instants
$t$ and $t+dt$ yields:
\begin{equation}
{dX\over dt}=\left\{ \begin{array}{ll}
	\Delta X\displaystyle{dm_{\rm t}\over dt}+m_{\rm t}\displaystyle{d\Delta X\over dt} 
		& \mbox{if $m>m_{\rm t}(t)$,} \smallskip\\
	\Delta X\displaystyle{dm_{\rm t}\over dt}-(1-m_{\rm t})\displaystyle{d\Delta X\over dt}\quad 
		& \mbox{if $m<m_{\rm t}(t+dt)$.}
		\end{array}
	\right.
\label{eq:dxdt}
\end{equation}
In this case, we assumed $m_{\rm t}(t)$ to be a decreasing function of time
(such as the plasma phase transition, which moves with time towards the center
of the planet). This has no consequence on the final result
however. The derivative $dX/dt$ is infinite between $m_{\rm t}(t)$ and
$m_{\rm t}(t+dt)$, but its integral over this mass interval is finite:
\begin{equation}
\int_{m_{\rm t}(t+dt)}^{m_{\rm t}(t)}-{dX\over dt}T\delta S dm=
T(m_{\rm t})\delta S(m_{\rm t})\Delta X{dm_{\rm t}\over dt}
\label{eq:intdx}
\end{equation}

Putting Eqs.~\ref{eq:ds}, \ref{eq:dxdt} and \ref{eq:intdx} into
Eq.\ref{eq:energy} yields: 
\samepage{
\begin{eqnarray}
{L\over M}&=&\int-T\left({\partial S\over \partial t}\right)_X dm 
	\nonumber \\
	& & +{d\Delta X\over dt}\left\{
\int_0^{m_{\rm t}} T \delta S dm -m_{\rm t}\int T \delta S dm
	\right\} \nonumber \\
	& & -\Delta X{dm_{\rm t}\over dt}\left\{
\int T \delta S dm - T(m_{\rm t})\delta S(m_{\rm t})
	\right\}
\label{eq:evol}
\end{eqnarray}
}
The first term on the right hand side of the equation corresponds to
homogeneous contraction and cooling of the planet. The second one,
proportional to $d\Delta X/dt$, is due to helium sedimentation. In our
case, $\Delta X$ is monotonically increasing with time. Moreover,
$T(m<m_{\rm t})>T(m>m_{\rm t})$ and $S(m<m_{\rm t})\sim S(m>m_{\rm
t})$ (except in very particular cases which are of little importance
here), so that this second term is positive. Helium sedimentation thus
provides an additional energy source. 

Finally, the third term,
proportional to $dm_{\rm t}/dt$, is caused by the displacement of the
phase separation level with time. Its sign is unknown: if the
transition between regions I and II follows the plasma phase
transition \citep{scvh95}, then $m_{\rm t}$ should
decrease with time. On the other hand, a more general phase diagram
could lead to either positive or negative $dm_{\rm t}/dt$. Finally the
bracketed term is generally negative, but only if $m_{\rm t}$ is not too
close to the center of the planet (more than $\sim 0.45$). If this is
verified, the whole third term is positive (i.e. provides energy) if
the transition moves towards the planetary center, and negative if
$m_{\rm t}$ gets closer to the surface.

\subsection{Numerical application}

We estimate quantitatively the significance of the different terms
constituting Eq.~\ref{eq:evol} using today's models of Jupiter and
Saturn. A more satisfactory approach would be the direct derivation of
the equation and the calculation of an evolution consistently taking
into account non-homogeneous effects. However, our simple approach is
justified, in view of the
uncertainties that remain on the hydrogen-helium phase diagram 
[\citet{klep91}, \citet{pfaff95};
see \citet{guill95} for a 
discussion] or on the presence of a first order molecular/metallic
hydrogen phase transition. 

We assume that the phase separation level is closely associated to the
molecular/metallic hydrogen transition and accordingly use the
derivation by \citet{scvh95} to predict the value of $m_{\rm t}$ and
its evolution with time. Its present value is $\sim 0.85$ in Jupiter
and $\sim 0.5$ in Saturn. This number can change depending on the
model, especially in the case of Saturn. 

Thus, we find:
\begin{eqnarray}
\int_0^{m_{\rm t}} T\delta S dm - m_{\rm t}\int T\delta S dm  & \simeq &
	\left\{\begin{array}{ll}
		5 \times 10^{11} {\rm \,erg\,g^{-1}}\ \quad & \mbox {for
	Jupiter} \\
		4 \times 10^{11} {\rm \,erg\,g^{-1}} \quad & \mbox {for
	Saturn}
	\end{array}\right. \\
\int T\delta S dm - T(m_{\rm t})\delta S(m_{\rm t})  & \simeq &
	\left\{\begin{array}{ll}
		2.5 \times 10^{12} {\rm \,erg\,g^{-1}} \quad & \mbox {for
	Jupiter} \\
		10^{11} {\rm \,erg\,g^{-1}} \quad & \mbox {for
	Saturn}
	\end{array}\right.
\end{eqnarray}
Furthermore, using homogeneous evolution calculations, we derive an
upper limit to the displacement of the transition with time:
\begin{eqnarray}
\left|{d m_{\rm t}\over dt}\right| < 2 \times 10^{-2}\,\rm Gyr^{-1},
\end{eqnarray}
this relation being valid for Jupiter and Saturn. We can
derive only an upper limit because any helium differentiation 
tends to slow the cooling and contraction of the planet. 

These numerical estimations can be put into Eq.~\ref{eq:evol} and
compared to the intrinsic luminosity per unit mass of Jupiter and Saturn, 
$5.7\times 10^{10}$ and $4.7\times 10^{10}\,\rm
erg\,g^{-1}\,Gyr^{-1}$, respectively. Static interior models in
agreement with the measured gravitational moments predict that $\Delta
X < 0.08$ in Jupiter and that $0 <\Delta X < 0.60$ in Saturn. 
The contribution due to the displacement of the transition level is 
of the order of $5\times 10^{10}\Delta X $ and $2\times 10^{9}\Delta X\,\rm
erg\,g^{-1}\,Gyr^{-1}$ in Jupiter and Saturn, respectively. 
It will therefore be neglected in this section.  The following
section presents an approximate treatment based on the
phase diagram of \citet{scvh95}.
In general, this effect may contribute as much as $\sim 5$\%
change to the calculated ages. 

The term proportional to $dm_{\rm t}/dt$ being thus ignored in
Eq.~\ref{eq:evol}, we derive the time delay due to a linear increase
of the mass fraction discontinuity $\Delta X$ over the time $\Delta
t$, by assuming that today's luminosity is entirely due to the
differentiation:
\begin{equation}
\Delta t \simeq \left\{\begin{array}{ll} 
	9.1\, \Delta X \quad {\rm Gyr} \qquad \mbox{for Jupiter,}\\
	8.3\, \Delta X \quad {\rm Gyr} \qquad \mbox{for Saturn.}
	\end{array}\right.
\end{equation}
Using model ages from homogeneous evolution calculations, and
reasonable uncertainties due to both the model atmospheres and to the
approximations inherent in our semi-analytical model, we can hence
constrain $\Delta X$. 

For Jupiter, calculations based on homogeneous evolution (see below)
give model ages between 3.6 and 5.1\,Gyr, the lower limit being
derived from models with a radiative zone and an interpolated equation
of state, and the upper limit from fully adiabatic models with a PPT
EOS. With a $\sim 0.3\,$Gyr uncertainty, we derive that, for Jupiter,
$-0.09 < \Delta X <  0.12$. This global constraint is not very useful
and has in fact to be linked to the various static models to
predict values of $\Delta X$. 
In the case of Saturn, homogeneous evolution models predict ages
between 2 and 3\,Gyr. It therefore appears that a more useful
constraint can be derived for this planet: $0.15 < \Delta X <
0.34$. 

\section{Core formation model}

\subsection{Helium differentiation}
\label{sec:hediff}

In this section we present an alternative approach in which
we estimate the heat
evolved during a single time step $\delta t$ in which
a dense component settles out from a hydrogen-helium mixture into
the core.  The dense component is taken to be helium, but
similar considerations apply to any constituent whose specific
entropy is much smaller than hydrogen's (as is true in general).
Thus we write a variant of Eq.~\ref{eq:energy} in the following form:

\begin{equation}
{\delta t}=-\left[\int dm T \delta S \right]/(L/M),
\label{eq:wbh1}
\end{equation}
where 
\begin{equation}
L=4 \pi R^2 \sigma (T_{\rm eff}^4-T_S^4),
\label{eq:wbh2}
\end{equation}
with $R$ the radius of the planet's photosphere and $T_S$ the
effective radiating temperature that the photosphere would have
if it were radiating only thermalized sunlight.  For homogeneous
evolution, $T_{\rm eff}$ and specific entropy $S$
(as parametrized by $T_{10}$) are uniquely related at each
point in the evolution, via the surface presented in Fig. 1.  We
treat inhomogeneous evolution by starting with an optimized model
of present Jupiter or Saturn with specified helium mass fraction
(relative to the total hydrogen-helium mass) $Y$ in the entire
hydrogen-rich portion of the planet.  The remaining helium or other
dense component is assumed to be incorporated in a central core.
In some sense, this model represents an upper limit on the heat release
(and concomitant prolongation of cooling) caused by a given amount of
helium separation, since the mass of low-specific-entropy material
displaced to the center of the planet is maximized.

We now consider evolution backward in time to a previous model in
which the helium incorporated in the core is restored to the
hydrogen-rich envelope, which then has a helium mass fraction equal to
$Y=Y_{\rm proto}=0.27$.  The time
step involved in this process is evaluated using Eq. (10).  At prior
times the evolution is homogeneous.

In the sample calculations presented here, we assume,
for Jupiter, that the planet evolves from a uniform $Y=0.27$
to a separated model at present with $Y=0.24$ in the entire
H part of the planet and pure He (corresponding to the depleted amount of
He) in a core at the center.  For Saturn, we assume the same scenario except
that the present model has $Y=0.20$ in the entire H part of the planet.

For both planets, the following effects occur between the
present-day model (shown with solid line in Fig. 2) and the previous
undifferentiated model.  First, the central temperature is {\it larger}
in the present-day model because of greater differentiation.  This
corresponds to the
energy retained in the form of a higher internal temperature mentioned
previously. Second, a mass element which is depleted in the denser
component (going forward in time) goes to a higher specific entropy,
because it is richer in H, while mass elements near the center of the
planet go to a lower specific entropy.  Because the latter are at a
higher temperature than the former, there is a net heat release going
forward in time.  The heat which is released to space
during this single time
step, i.e. total heat release less heat stored via heating of the
interior due to core formation,
is $1.9 \times 10^{11} {\rm \,erg\,g^{-1}}$
for Jupiter, and $1.8 \times 10^{11} {\rm \,erg\,g^{-1}}$ for Saturn.
Similar values are found from Eqs. (5) and (6), when assuming $\Delta
X\sim 0.8$ and $m_{\rm t}\sim 0.04$ for Jupiter and $m_{\rm t}\sim
0.09$ for Saturn (where in this case $m_{\rm t}$ denotes the mass fraction
of the pure helium core).

\subsection{Variable and constant solar heating}
\label{sec:solheat}

Figure 3 shows the evolution of Jupiter and Saturn for three assumptions
about solar heating.  The crosses show the evolution of isolated
objects with $T_S=0$.  The open circles show evolution with
$T_S$ held constant at its present value for each planet, while the
dots show evolution with a variable $T_S$ computed on the assumption that
the solar luminosity has increased roughly linearly with time from
an initial value of about 72\% of the present luminosity.

\begin{figure*}

\epsfxsize=15cm \epsfbox{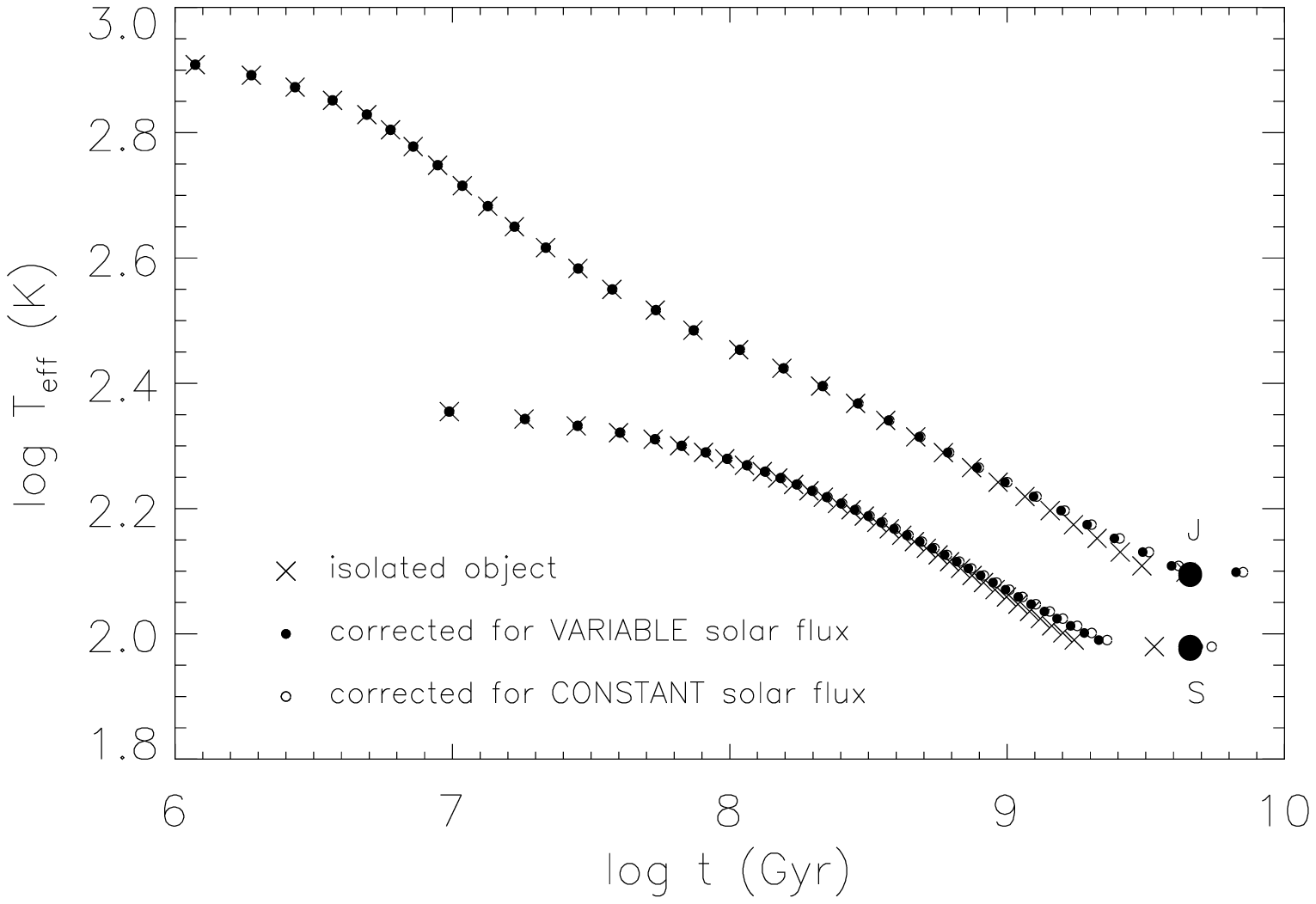}
  \caption[]{\label{fig3:solheat}
Variation of $T_{\rm eff}$ vs. $t$ for homogeneous (solar-composition)
Jupiter and Saturn; large dots show present values.  The final time
steps illustrate the effect of He differentiation and are shown in
expanded scale in the next figure.}
\end{figure*}

This calculation assumes than Jupiter's Bond albedo has
remained constant in time at its present value (0.343).  However
it is likely that during the time water clouds first condensed
in the planet's atmosphere (near $T_{\rm eff}=400$ K), the Bond albedo
was substantially higher for a time until the clouds moved
lower in the atmosphere with decreasing $T_{\rm eff}$.  At earlier times
still the planet would be free of clouds of abundant species,
yet the Bond albedo would be similar to the current time.  \citet{marl99}
discuss Bond albedos for Jupiter-like models with
and without clouds.

Figure 4 shows an expanded view of the final stages of evolution of
Jupiter and Saturn.  The vertical error bars show the present values
for the two planets.
The heavy horizontal error bar shows the
prolongation of ages for the $Y_{\rm proto}-Y$
assumed in this section. However, it
assumes the linear differentiation with time of the previous section, and
that the helium separates into the metallic hydrogen region but not the
very center of the planet.

\begin{figure*}

\epsfxsize=15cm \epsfbox{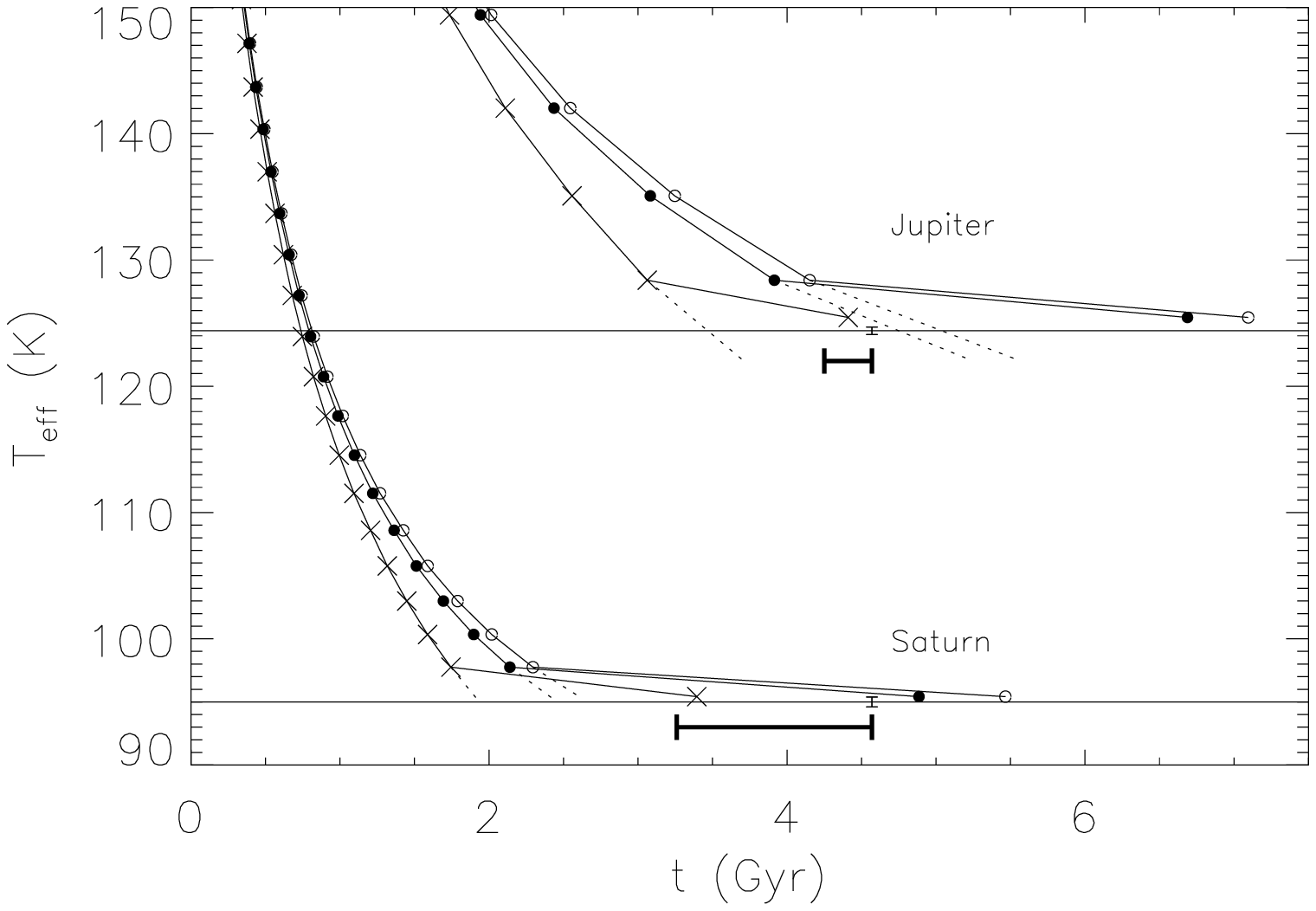}
  \caption[]{\label{fig4:solheat}
Expanded view of the final stages of evolution of Jupiter and Saturn,
assuming that He differentiation occurs in the final time step.  Dashed
curves show evolution without He differentiation. Heavy horizontal error
bars show prolongation of evolution for the same atmospheric helium
depletion assumed here, but using the theory of Section 2.}
\end{figure*}

\section{Summary}

Our conclusions are as follows.  

(a) The atmospheric boundary conditions are
reasonably well understood (see Fig. 1) and are no longer
a source of major uncertainty in the cooling theory for Jupiter
and Saturn.  With these boundary conditions and the assumption
of homogeneous evolution,
Jupiter cools to its present $T_{\rm eff}$ in 3.6 to 5.2 Gyr,
while Saturn cools to its present $T_{\rm eff}$ in approximately 2
to 3 Gyr.
Therefore differentiation is needed to account for Saturn's present
luminosity, but not necessarily for Jupiter's.

(b)  Solar heating prolongs the cooling time of Jupiter by about
1 Gyr and the cooling time of Saturn by about 0.5 Gyr.  Allowance
for lower solar heating in the past reduces the effect by $\sim 0.1$ Gyr
for both planets, a very minor effect.
Likewise variation in the planetary Bond albedo with time
will have comparably small effects on the evolution.

(c) We have considered two limiting models for He differentiation in Jupiter.
The model presented in Section 3 shows that 
a pure helium
core cannot have formed in Jupiter and result in the presently
observed atmospheric $Y$.
Such a recent differentiation in Jupiter,
corresponding to a reduction of the atmospheric $Y$ by about 0.03
(the Voyager result) would indeed prolong the evolution by about
another 2 Gyr.
In contrast,
the model presented in Section 2 predicts that the difference
between the primordial helium mass fraction $Y_{\rm proto}$ and
the present atmospheric $Y$ should lie in the range
$-0.08 < Y_{\rm proto}-Y < 0.10$.
Note that a value of zero for
$Y_{\rm proto}-Y$ corresponds to no
differentiation, whereas a negative value indicates an unlikely upward
transport of any element other than hydrogen.
However models for which $Y_{\rm proto}-Y>0.07$ cannot reproduce the
planet's gravitational field \citep{guill99}.
Thus $0.20 < Y < 0.27$. This is higher than
the Voyager value $Y=0.18$ \citep{go89} but it includes the new Galileo
value $Y=0.238$ \citep{vonzahn98}.

(d) We have considered two limiting models for He differentiation in Saturn.
The model presented in Section 3 shows that
recent differentiation of helium in Saturn, corresponding
to a reduction of the atmospheric $Y$ by about 0.07
would
prolong the evolution by about another 2 Gyr, bringing agreement
between the observed age and $T_{\rm eff}$.
The model presented in Section 2 predicts that
$0.06 < Y_{\rm proto}-Y < 0.14$, and hence $0.13 < Y < 0.21$.
Neither limiting model includes the Voyager value $Y=0.06\pm
0.05$. The case for a higher atmospheric $Y$ in Saturn is further
strengthened by static models, the enrichment in heavy elements
required to fit the planetary gravitational field being incompatible
with the observed methane abundance for $Y<0.11$ \citep{guill99}.

(e) The models presented must both be corrected for the fact that the
energy that results from differentiation is not added entirely at the
end of the evolution. Should helium differentiation occur very early in
the evolution (a very unlikely assumption), the subsequent time delay
would be negligible compared to the present calculations. In most
likely cases, we estimate that it yields time delays overestimated
by $\sim 5$\% for Jupiter, and by $\sim 10-30$\% for Saturn.

(f) The most precise
determination of the Jovian helium abundance in the outer atmosphere is
that of the Galileo Helium Abundance detector, and our results are in
excellent agreement with that measurement. The Voyager-determined Jovian
helium abundance lies below both, and indeed for Saturn we also predict a
helium abundance larger than that determined by Voyager. Both the very
precise Galileo HAD determination and that of the Galileo mass spectrometer
\citep{niem98}, which is fully consistent with but less precise
than the HAD value, give us confidence in our model results. In fact, we
suggest that a reexamination of the Voyager helium determinations for both
planets may be order, specifically with the aim of assessing whether the
initial analyses has errors that systematically lowered the helium
abundances. This exercise is not an academic one, because the Cassini
mission enroute to Saturn will not be capable of providing in situ
measurements of the helium abundance. The Galileo Jupiter results combined
with ours raise a potential concern that the Cassini remote sensing
measurements to determine helium in Saturn, using CIRS and radio
occultations, might be subject to similar systematic errors.

It must be stressed that these are still preliminary models.
A more detailed study interfaced to a complete physical picture of the H-He
phase diagram is in preparation.  We should also note that in either limiting
model, the extra heat which is liberated could come from the differentiation
of some other dense component beside helium.  Thus an observational result
indicating a primordial solar value for the atmospheric helium abundance
would not necessarily preclude extension of the cooling time via
differentiation of another abundant component.

\begin{acknowledgements}

Research supported by the National Aeronautics and Space Administration
(Origins Program),
the National Science Foundation, and the ``Groupe de Recherche Structure
Interne des Etoiles et des Plan\`etes G\'eantes''.

\end{acknowledgements}


\begin{thebibliography}{}    

\bibitem[Burrows et~al.(1997)]{burr97}
Burrows, A., Marley, M., Hubbard, W.~B., Lunine,, J.~I.,
Guillot, T., Saumon, D., Freedman, R., Sudarsky, D., and Sharp, C.,
A non-gray theory of extrasolar giant planets and
brown dwarfs,
{\em Astrophys. J.}, {\em 491}, 856--875, 1997.

\bibitem[Collins et~al.(1998)]{collins98}
Collins, G. W., Da Silva, L. B., Celliers, P., Gold, D. M., Foord, M. E.,
Wallace, R. J., Ng, A., Weber, S. V., Budil, K. S., and Cauble, R.,
Measurements of the equation of state of deuterium at the fluid
insulator-metal transition, {\em Science}, {\em 281}, 1178--1181, 1998.

\bibitem[Gautier and Owen(1989)]{go89}
Gautier, D., and Owen, T., 1989.  The composition of outer planet atmospheres.
In: Atreya, S.~K., Pollack, J.~B., and Matthews, M.~S. (Eds.),
Origin and Evolution of Planetary and Satellite Atmospheres.
The University of Arizona Press, Tucson, Arizona and London. 

\bibitem[Guillot et~al.(1995)]{guill95}
Guillot, T., Chabrier, G., Gautier, D., and Morel, P,, 
Effect of radiative transport on the evolution of Jupiter and Saturn,
{\em Astrophys. J.}, {\em 450}, 463--472, 1995.

\bibitem[Guillot(1999)]{guill99}
Guillot, T., A comparison of the interiors of Jupiter and Saturn,
submitted to {\em Planet. Space Sci.}, 1999.

\bibitem[Hubbard(1977)]{wbh77}
Hubbard, W.~B., The jovian surface condition and cooling rate,
{\em Icarus}, {\em 30}, 305--310, 1977.

\bibitem[Hubbard et~al.(1997)]{wbh97}
Hubbard, W.~B., Guillot, T., Lunine, J.~I.,
Burrows, A., Saumon, D., Marley, M.~S., and Freedman, R.~S.,
Liquid metallic hydrogen and the structure of
brown dwarfs and giant planets,
{\em Physics of Plasmas}, {\em 4}, 2011--2015, 1997.

\bibitem[Klepeis et~al.(1991)]{klep91}
Klepeis, J.~E., Schafer, K.~J., Barbee, T.~W., III, and Ross, M.,
Hydrogen-helium mixtures at megabar pressures: Implications for
Jupiter and Saturn,
{\em Science}, {\em 254}, 986--989, 1991.

\bibitem[Marley et~al.(1999)]{marl99}
Marley, M.~S., Gelino, C., Stephens, D.,
Lunine, J., and Freedman, R.,
Reflected spectra and albedos of extrasolar giant planets I: Clear and cloudy
atmospheres, {\em Astrophys. J.}, {\em 513}, in press, 1999.

\bibitem[Niemann et~al.(1998)]{niem98}
Niemann, H.~B., Atreya, S.~K., Carignan, G.~R., Donahue, T.~M., Haberman, J.~A.,
Harpold, D.~N., Hartle, R.~E., Hunten, D.~M., Kasprzak, W.~T., Mahaffy, P.~R.,
Owen, T.~C., and Way, S.~H., The composition of the Jovian atmosphere as
determined by the Galileo probe mass spectrometer,
{\em J. Geophys. Res.}, {\em 103},
22831--22845, 1998. 

\bibitem[Pfaffenzeller et~al.(1995)]{pfaff95}
Pfaffenzeller, O., Hohl, D., Ballone, P., 
Miscibility of hydrogen and helium under astrophysical conditions,
{\em Phys. Rev. Lett.}, {\em 74}, 2599--2602, 1995.

\bibitem[Saumon, Chabrier, and Van Horn(1995)]{scvh95}
Saumon, D., Chabrier, G., and Van Horn, H.~M.,
An equation of state for low-mass stars and giant planets,
{\em Astrophys. J. Suppl.}, {\em 99}, 713--741, 1995.

\bibitem[Saumon et~al.(1992)]{saumon92}
Saumon, D., Hubbard, W.~B.,
Chabrier, G., and Van Horn, H.~M.,
The role of the molecular-metallic transition of hydrogen in the
evolution of Jupiter, Saturn, and brown dwarfs,
{\em Astrophys. J.}, {\em 391}, 827--831, 1992.

\bibitem[Stevenson and Salpeter(1977)]{ss77}
Stevenson, D.~J., and Salpeter, E.~E.,
The dynamics and helium distribution properties
for hydrogen-helium fluid planets,
{\em Astrophys. J. Suppl.}, {\em 35}, 239--261, 1977.

\bibitem[von Zahn, Hunten, and Lehmacher(1998)]{vonzahn98}
von Zahn, U., Hunten, D.~M., and Lehmacher, G., Helium in Jupiter's
atmosphere: Results from the Galileo probe helium interferometer experiment,
{\em J. Geophys. Res.}, {\em 103}, 22815--22829, 1998.


\end{thebibliography}

\end{document}